\documentclass[fleqn,10pt]{wlscirep}

\usepackage{framed,multirow}
\usepackage{amssymb}
\usepackage{latexsym}
\usepackage{url}
\usepackage{xcolor}
\usepackage{hyperref}
\usepackage{cite}
\usepackage{amsmath}
\usepackage{algorithmic}
\usepackage{booktabs}
\usepackage[T1]{fontenc}
\usepackage[utf8]{inputenc}
\usepackage[euler]{textgreek}

\newcommand{\rev}[1]{\textcolor{black}{#1}}

\title{DiagSet: a dataset for prostate cancer histopathological image classification}

\author[1,2,*]{Michał Koziarski}
\author[1,2]{Bogusław Cyganek}
\author[2]{Przemysław Niedziela}
\author[1]{Bogusław Olborski}
\author[1]{Zbigniew Antosz}
\author[1]{Marcin Żydak}
\author[1,2]{Bogdan Kwolek}
\author[1]{Paweł Wasowicz}
\author[2]{Andrzej Bukała}
\author[1,3]{Jakub Swadźba}
\author[1]{Piotr Sitkowski}
\affil[1]{Diagnostyka Consilio Sp. z o.o., Ul. Kosynierów Gdyńskich 61a, 93-357 Łódź, Poland}
\affil[2]{AGH University of Science and Technology, Al. Mickiewicza 30, 30-059 Kraków, Poland}
\affil[3]{Andrzej Frycz Modrzewski Krakow University, Gustawa Herlinga-Grudzińskiego 1, 30-705 Kraków, Poland}

\affil[*]{michal.koziarski@gmail.com}

\begin{abstract}
Cancer diseases constitute one of the most significant societal challenges. In this paper, we introduce a novel histopathological dataset for prostate cancer detection. The proposed dataset, consisting of over 2.6 million tissue patches extracted from 430 fully annotated scans, 4675 scans with assigned binary diagnoses, and 46 scans with diagnoses independently provided by a group of histopathologists can be found at \url{https://github.com/michalkoziarski/DiagSet}. Furthermore, we propose a machine learning framework for detection of cancerous tissue regions and prediction of scan-level diagnosis, utilizing thresholding to abstain from the decision in uncertain cases. The proposed approach, composed of ensembles of deep neural networks operating on the histopathological scans at different scales, achieves 94.6{\%} accuracy in patch-level recognition and is compared in a scan-level diagnosis with 9 human histopathologists showing high statistical agreement.
\end{abstract}

\begin{document}

\flushbottom

\maketitle

\thispagestyle{empty}

\section*{Introduction}

In highly developed countries, prostate cancer is the second most common cause of death in men after lung cancer. Prostate cancer is one of the most common malignant neoplasms in men. Treatment method choice depends mainly on the clinical stage and malignancy determined by a specialist histopathologist according to the Gleason scale \cite{Albelda1993,Konig2004,Gleason1966} and the ISUP classification 2014 by the ISUP grade group \cite{Epstein2016}. However, the number of professional doctors is limited and continues to decline compared to social needs. A way out of this situation is using modern technologies based on deep learning and the necessary repositories of labeled data to train them. 

In this paper, we present and share a new set of histopathological data, called DiagSet, containing the annotated regions of prostate tissues in the whole slide imaging (WSI) scans characterized by different Gleason degrees. We also present the structure and the results of the operation of the individual \rev{deep convolutional neural networks (CNNs) and vision transformers (ViTs) as well as an ensemble of the deep convolutional neural networks} appropriately trained for the classification of histopathological images of the prostate WSI. First, AlexNet \cite{krizhevsky2012imagenet}, VGG16, VGG19 \cite{simonyan2014very}, ResNet50 \cite{he2016deep}, \rev{Inception V3 \cite{szegedy2016rethinking}, and ViT-B/32 \cite{transformer_16x16}} networks were trained and their performance analysed. Second, these \rev{convolutional} networks were used again, but this time cooperating in an ensemble, wherein each classifier was trained with images of a different magnification factor, after which the individual probabilities returned by each network were combined to produce the final result. Then few rules have been developed and used to produce the binary diagnosis. The best binary classification ratio obtained with the ensemble architecture is 94.58{\%}. An experiment was then conducted in which nine volunteer professional histopathology doctors participated, making individual diagnoses on a set of forty-six anonymous WSI scans. Then the correlations of their responses were computed, as well as correlations of their responses with respect to the machine given diagnosis by the proposed method. The conducted statistical analysis showed that, except for one WSI scan, the examination correlation of the machine response is within the correlation range obtained between the participating doctors.  





Regarding the literature and recent works, the classification of histopathological scans can be realized by various types of CNN \cite{krizhevsky2012imagenet,simonyan2014very,szegedy2016rethinking,Chetan2019,Rczkowski2019ARAAR}, recurrent neural networks (RNN), as well as autoencoders (AE) \cite{he2016deep}, generative adversarial networks (GAN) \cite{Goodfellow-et-al-2016}, Transformers \cite{transformer_16x16,swin_transformer}, or more complex systems composed of these, respectively. Also, different histopathological features can be applied, such as entire cells, nuclei, glands, tissue texture, or a combination of these \cite{Spanhol:2016,WAN201734,Pan2017AccurateSO}. CNN models in cancer diagnosis histopathology are presented e.g. in the papers by Chetan et al. \cite{Chetan2019}, Litjens et al. \cite{LITJENS201760}, as well as Janowczyk and Madabhushi \cite{Janowczyk2006}. It was shown that the modern data-oriented approaches with CNN outperform the previously developed methods based on expert proposed hand-crafted features and models. 
The patch-based cancer classification of the WSI scans was employed by many researchers; for example, Litjens et al. proposed the CNN for the prostate and breast cancer diagnosis from the H\&E scans \cite{Litjens2016}, while Vandenberghe et al. for the breast cancer \cite{Vandenberghe2017}. In all such systems, the vital part is a preparation of the training datasets with proper patch labeling. 
In the case of prostate cancer, which we are concerned about mainly in this paper, a degree of prostatic carcinoma is described with the well-established Gleason scale \cite{Gleason1966,Epstein2005}. Therefore, in the case of the supervised classification of prostate cancer, the most natural labeling is just based on the Gleason scores. In this work, we also follow this strategy.
For example, Bulten et al. propose grading and prostate cancer detection based on the UNet segmentation to the growth of the Gleason patterns, which are followed by the subsequent cancer grading \cite{Bulten2019}. Campanella et al. also proposed a system for prostate cancer detection in WSI \cite{campanella2018terabytescale}. Arvaniti et al. discuss the problem of deep multiple instance learning for the classification and localization of prostate cancer \cite{Arvaniti2018}.
On the other hand, the classification of the prostate tissues into tumor vs. non-tumor based on convolutional adversarial autoencoders was proposed by Bulten and Litjens \cite{Bulten2018}. For this purpose, the WSI dataset from the Radboud University Medical Center has been used, which scans are hematoxylin and eosin (H\&E) stained. 
Interestingly, Ren et al. proposed unsupervised training of the Siamese neural network for the prostate WSI patch based classification \cite{Ren2018}.

Research on the development and validation of a deep learning algorithm for improving Gleason scoring of prostate cancer has been performed by Nagpal et al. \cite{Nagpal2019}. Because the Gleason scoring among pathologists is highly subjective and suffers from inter and intra-observer variability, with the discordance ratio reported to be in the range 30\%-50\%, Nagpal et al. in investigate performance of the deep neural network, for providing an automatic and reproducible method for feature extraction. 

Nir et al. conducted extensive research into the automatic grading of prostate cancer in digitized histopathology images \cite{Nir2018}. Extensive experiments with various classifiers, such as linear discriminant analysis (LDA), support vector machines (SVM), logistic regression (LR), and random forests (RF), operating with a broad set of hand-crafted features, as well as deep neural networks, were carried on. Interestingly, the best results in terms of accuracy and overall agreement were obtained by LR with the hand crafted features, whereas the worst were observed with SVM. Also, interestingly DNN performed well, but not the best, due to the insufficient number of training data and probably not sufficiently deep architecture, as concluded by the authors. All methods as mentioned above, based on deep structures, such as CNN, but also Vision Transformers (ViT), depend entirely on training datasets, mainly their size and quality of labeling. Therefore, in this work, we want to fill this gap by providing a new set of prostate histopathology data, as well as baseline implementations of various variants of deep classifiers.

\rev{Recently, several new databases with annotated histopathological images and their classification algorithms have been proposed, such as the PANDA challenge by Bulten et al. \cite{Bulten2022}. Nevertheless, the work described in this article provides yet another database of valuable histopathological data, and is complementary to the database from the PANDA project.} 

\rev{An interesting system for automatic end-to-end histology prostate grading and cribriform pattern detection was proposed by Silva-Rodríguez et al. \cite{Silva2020}. Their work is mostly focused upon automatic detection of individual cribriform patterns belonging to Gleason grade 4. In the future, both the data presented there and the methodology may be incorporated into our computing environment, e.g. in the form of ensemble methods.}

\rev{Finally, in order to assist in research and collect the best ideas and solutions in the field of automated Gleason grading in computational pathology, the Automated Gleason Grading Challenge (AGGC) 2022 competition was organized \cite{AGGC2022}. The main goal was to develop algorithms for Gleason patterns identification in the H\&E stained WSI. The AGGC22 dataset was obtained from National University Hospital, Singapore, with all annotations done by experienced pathologists. In order to to assess the variations caused by the digitalization process, all specimens are scanned by multiple scanners. The idea of AGGC22 is similar to our research. However, all other features are different, since we used different specimens, the annotations were done by other expert pathologists, as well as we used a different scanner by Hamamatsu. The best performing methods from AGGC 2022 are still to be published. Hence, in the future it will be interesting to check the best performing AGGC22 method with our DiagSet.}

\section*{Methods}

\subsection*{Dataset}
\label{sec:dataset}

\noindent\textbf{Data acquisition.} We created a fully anonymized dataset of WSI scans that contained no patient data. The dataset was constructed based on the already scanned material, gathered during standard laboratory diagnosis, which was anonymized and later annotated by a group of histopathologists. No medical experiments were done to gather the data. Because of the above, the data gathering process required no institutional approval or informed consent according to national regulations in Poland, specifically the \textit{Patients' Rrights and Patient Ombudsman Act} from 6th November 2008, and was carried out in accordance with relevant guidelines and regulations.

For preparation of dataset randomly selected microscopic specimens of biopsy specimens, i.e., sections from prostate tumors diagnosed with adenocarcinoma of the prostate, were subjected to research and experiments. Microscopic slides were made in the classical formalin-paraffin technique in the histopathology laboratory. Prostate biopsies, after fixation in 10\% formalin (10\% buffered formalin, 4\% formaldehyde content, manufacturer: Alpinus Chemia), were embedded in paraffin blocks. Subsequently, in the course of cutting (MICROM HM355S microtome), preparations with a thickness of 5-7 micrometers were obtained, which were stained with hematoxylin and eosin (H\&E). The diagnostic assessment was performed by doctors pathomorphologists employed at Diagnostyka Consilio.

All scans were acquired with the Hamamatsu C12000-22 digital slide scanner. The scanner uses time delay integration (TDI) scanning method. The magnification (objective lens) was $40\times$. All slides were scanned in one $z$-stack layer with dynamic pre-focus and pre-focus map.  
Before scanning, slides were inspected for overhanging labels and traces of the felt tip pen. They where then wiped with soft cloth to remove loose debris, water spots, or fingerprints from the upper and lower surface. For difficult slides alcohol solution were used.
Often, when tissue of prostate needle biopsy was narrower than 0.5 mm, it was omitted in the scanning area. It happened once in every 300 cases of the prostate needle biopsy slides. Sometimes, if parts of the slides were blurry, it was caused by the folded or rugged surface of a tissue.

The slides were scanned in unattended mode using the feeder for 320 slides. The area of the scan, as well as the focal points, were set automatically. The median scan time per slide at $40\times$ equivalent resolution (0.25 \textmu m/pixel) was 3 minutes. The average file size was 1.2 gigabytes (GB). To recognize the tissue type, we use barcode recognition and then query the external slide database. 
All images were stored on the disk array in the NDP format. We used the Slide distribution and management software NDP.server3. The NDP.Server software API was used to access the slide images, annotations, and detailed slide data (magnification, scan info, barcode label, etc.). The JPG slide images were acquired using the NDP tiles server API.

\noindent\textbf{Grading protocol.} As already mentioned, the assessment of prostate cancer is performed according to the Gleason scale \cite{Albelda1993,Konig2004,Gleason1966}, as well as the newer ISUP grades \cite{Epstein2016}. The values of the former run from 1 (mild) up to 5 (highly malignant). Then, the pathologist expert provides an assessment/diagnosis based on the sum of the points (Gleason score; GS) allocated to the most diverse zones within the cancer area in the tissue/specimen tested. First, the score for \textit{the dominant} feature in the examined tissue is given, and then the score for \textit{the second largest} feature. The result is given as the sum of these scores. Hence, the final GS sum can be from 2 (i.e., 1+1) to 10 (5+5). On this basis, a simple cancer grading system in the preparation was defined:
GS sum 2-4 - low grade I tumor malignancy,
5-7 - moderate II degree of tumor aggression,
8-10 - very high III degree of tumor aggression.

On the other hand, the ISUP grades method \cite{Epstein2016} assumes five mischief groups that combine different GS grades as follows:
group 1 GS sum 3+3=6,
group 2 is 3+4=7,
group 3 is 4+3=7,
group 4 all GS combinations giving a sum of 8,
group 5 all GS combinations giving a sum of 9. However, in the ISUP scale Gleason grades 1 and 2 are omitted.

In our work, three pathologist experts evaluated the scans on original and archival microscopic slides without any prior diagnosis. 
Each of the outlined regions on a scan was assigned a single label out of 9 possible classes: scan background (BG), tissue background (T), normal, healthy tissue (N), acquisition artifact (A), or one of the 1-5 Gleason grades (R1-R5). 

\begin{figure*}
\centering
\includegraphics[width=\textwidth]{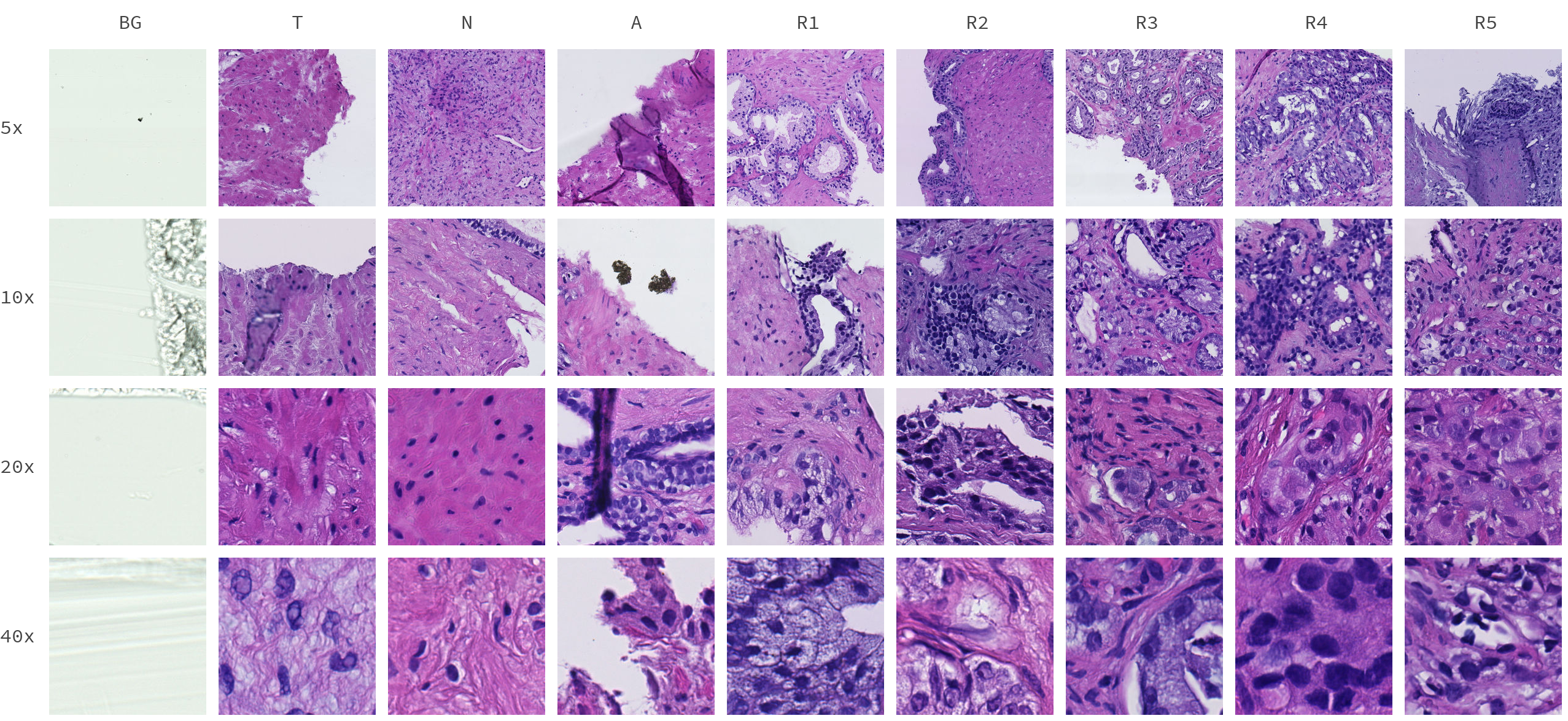}
\caption{Randomly selected samples of image patches from DiagSet-A, extracted at different magnifications (rows) and containing different classes of tissue (columns).}
\label{fig:diagset-samples}
\end{figure*}

\noindent\textbf{DiagSet-A.} The first part of the proposed dataset, DiagSet-A, consists of small image patches extracted from the underlying WSI scans, with labels assigned based on the annotation made by human histopathologists. Patches with a size of $256\times256$ were extracted from the scans with a stride of 128, at 4 different magnification levels: $40\times$, $20\times$, $10\times$ and $5\times$. 
Samples for each class and magnification level are presented in Figure~\ref{fig:diagset-samples}.

During the labeling process, a histopathologist annotated larger WSI regions as belonging to one of the defined classes. Due to the nature of the labeling process, some patches can be covered by annotations only partially, or contain multiple overlapping annotations. To translate these annotations to labels on the patch level the following procedure was used: on the highest magnification level, that is $40\times$, a label was assigned if and only if only a single class annotation with overlap ratio equal to or higher than 0.75 was present. In this case, that annotation label was assigned as a class associated with a given patch. If either none of the classes overlapped the patch at a specified ratio, or multiple contradictory labels were present, the patch was not assigned any class. Secondly, on lower magnification levels, that is $20\times$, $10\times$, or $5\times$, a patch was first divided into smaller $40\times$ patches (4 in case of $20\times$ magnification, 16 in case of $10\times$, and 64 in case of $5\times$). Each $40\times$-level patch was assigned a label according to the previously described procedure. Finally, the most severe of the $40\times$-level labels were assigned as a final label for the lower magnification patch. For instance, if given a $20\times$-level patch could be divided into one $40\times$-level patch with a label N, two $40\times$-level patches with an R3 label, and one $40\times$-level patch with an R4 label, the R4 label would be assigned to the $20\times$-level patch.

Due to the length of the annotation process, as well as the time required to train described machine learning models on large quantities of data, and for the sake of the experimental study described in the later part of this paper, DiagSet-A was divided into two parts. Firstly, DiagSet-A.1, consists of 238 WSI scans annotated by the histopathologists. DiagSet-A.1 was used in the preliminary investigation of architecture and ensembling choice. Secondly, DiagSet-A.2, which when compared to DiagSet-A.1, consists of 190 additional training scans, as well as a single additional validation and test scan. Importantly, DiagSet-A.2 also introduced an additional class, BG, which was not initially present in the dataset. DiagSet-A.2 was used during the final evaluation of the proposed approach, and can be treated as a final version of the dataset. Detailed number of scans and patches extracted for both versions of the dataset are presented in Table~\ref{table:diagset-size}, whereas their class distribution is presented in Table~\ref{table:diagset-distrib}.


\begin{table*}
\small
\caption{Class distribution of DiagSet-A.}
\label{table:diagset-distrib}
\centering
\begin{tabular}{lllllllll}
\toprule
\textbf{BG} & \textbf{T} & \textbf{N} & \textbf{A} & \textbf{R1} & \textbf{R2} & \textbf{R3} & \textbf{R4} & \textbf{R5} \\
\midrule
6.6\% & 23.8\% & 35.3\% & 8.3\% & 0.7\% & 0.7\% & 6.1\% & 15.4\% & 3.0\% \\
\bottomrule
\end{tabular}
\end{table*}


\begin{table*}
\small
\caption{Detailed number of scans and extracted patches in DiagSet-A.}
\label{table:diagset-size}
\centering
\begin{tabular}{lllll}
\toprule
\textbf{Partition} & \textbf{\# of scans} & \textbf{Magnification} & \textbf{\# of patches} \\
\midrule
train & 346 & $40\times$ & 1,250,661 \\
&  & $20\times$ & 398,201 \\
&  & $10\times$ & 132,882 \\
&  & $5\times$ & 48,782 \\
\cmidrule{1-4}
validation & 42 & $40\times$ & 245,441 \\
&  & $20\times$ & 77,889 \\
&  & $10\times$ & 25,294 \\
&  & $5\times$ & 8,977 \\
\cmidrule{1-4}
test & 42 & $40\times$ & 284,032 \\
&  & $20\times$ & 91,125 \\
&  & $10\times$ & 30,086 \\
&  & $5\times$ & 10,836 \\
\cmidrule{2-4}
& 430 &  & 2,604,206 \\
\bottomrule
\end{tabular}
\end{table*}

\noindent\textbf{DiagSet-B.} The second part of the presented dataset, DiagSet-B, consists of 4675 WSI scans with a singly binary diagnosis denoting either a presence of cancerous tissue on the scan (C) or lack thereof (NC), with 2090 scans belonging to the C class, and 2585 belonging to the NC class, respectively. These scans were extracted from the archive of past treatments, with labels assigned based on the text of the diagnosis given by a human histopathologist. Label assignment based on the text of the diagnosis was conducted manually. It should be noted that compared to the DiagSet-A, which was annotated solely based on the underlying WSI scan, the diagnoses used in DiagSet-B were given in a normal course of treatment and were potentially based on additional medical data, such as the results of the immunohistochemistry examination (IHC). According to the current procedure, each preparation was assessed independently by two pathologists.

\noindent\textbf{DiagSet-C.} The third part of the presented dataset, DiagSet-C, consists of 46 WSI scans with a global diagnosis given independently by a larger number of 9 human histopathologists. Unlike DiagSet-B, while labeling scans in DiagSet-C, histopathologists were asked to assign each scan one of the three possible labels: containing cancerous tissue (C), not containing cancerous tissue (NC), or uncertain and requiring further medical examination (IHC). Compared to DiagSet-B, including IHC in the set of possible labels more closely resembles the actual process of the histopathological diagnosis, in which a WSI scan is often insufficient to make a decision. Furthermore, aggregating the diagnoses of several medical practitioners allows us to evaluate the agreement within the population of histopathologists, as will be discussed.



\subsection*{Histopathological image recognition}
\label{sec:meth}

When dealing with histopathological image recognition models, operating on WSI scans, two main tasks can be distinguished. First of all, recognition of scan regions containing cancerous tissue, either on a binary cancerous/non-cancerous level, or a more fine-grained recognition of cancer types, such as the Gleason grades. Secondly, the prediction of an overall diagnosis for the whole scan, possibly based on the previously recognized cancerous regions. In this paper, we consider a methodology dealing with both of these steps, and relying on the fragmentation of the whole WSI scan into small image patches, that can then be treated as individual images in the image classification task. The proposed methodology consists of the following steps:

\noindent\textbf{Valid tissue segmentation:} in the initial step of the proposed machine learning pipeline, we conducted a preprocessing in the form of valid tissue segmentation, with the aim to reduce the computational overhead associated with the classification of a large number of individual patches. To this end we extract an image of a whole WSI scan downsampled with the factor of 8 and, using a fully convolutional neural network, perform a supervised image segmentation, with the goal to predict a binary mask containing information on whether any given pixel contains valid tissue (that is, not scan background or acquisition artifact). Since typically, a majority of WSI scan consists of the scan background, this operation can have a substantial impact on the computational overhead. In this step, we used a variant of fully-convolutional VDSR network \cite{kim2016accurate}, which consisted of 10 convolutional layers with 64 $3\times3$ filters each, trained on $21\times21$ image patches for 600 epochs using Adam optimizer with learning rate equal to 0.0001 and weight decay equal to 0.0001. \rev{Nevertheless, other image segmentation methods, such as the Otsu's one, can be used as well \cite{Otsu}.}

\noindent\textbf{Single-model patch recognition:} in the second step of the proposed pipeline, we performed image classification using small scan patches, extracted from the original WSI scan. The goal of this step was to produce probability maps for a given scan, indicating the likelihood that tissue at a given spatial position belongs to one of the predefined classes. To this end, we divided the scan into $224\times224$ patches and independently classified every one using a previously trained convolutional neural network. Importantly, we conducted this procedure for several neural architectures, as well as several magnification factors, to compute a collection of probability maps for each of the model/magnification combinations.

It is worth noting that such classification of small WSI patches is equivalent to rough image segmentation, with many notable approaches for this problem already existing, such as U-Net networks. Nevertheless, in this paper, we decided to formulate the problem as a patch classification task due to two reasons: 1) we suspected that data imbalance could pose a significant challenge given the dataset characteristic, mainly the class distribution, and a larger body of methods for dealing with class imbalance within the classification framework already exists, and 2) we decided that prediction granularity with sufficiently small patches, such as $224\times224$, will be sufficient for our purposes.

Throughout the conducted experimental study we considered several notable architectures of the convolutional neural networks \rev{and vision transformers} proposed in recent years: AlexNet \cite{krizhevsky2012imagenet}, VGG16 and VGG19 \cite{simonyan2014very}, ResNet50 \cite{he2016deep}, \rev{InceptionV3 \cite{szegedy2016rethinking} and ViT-B/32 \cite{transformer_16x16}}. All of the models were trained for 50 epochs using the SGD optimizer with an initial learning rate equal to 0.0001, decayed after every 20 epochs with a rate of 0.1, and batch size equal to 32. All of the models were regularized using weight decay equal to 0.0005. Additionally, AlexNet and both VGG models used a dropout equal to 0.5. To augment the data, during training, we cropped random $224\times224$ patches from the original $256\times256$ images, and afterward applied random horizontal flip and rotation by a random multiple of a 90-degree angle. During the evaluation we instead used a central cropping to obtain the same patch size. In both cases, \rev{for CNNs} input images were preprocessed by subtracting the ImageNet \cite{deng2009imagenet} image mean, that is, a tuple (123.680, 116.779, 103.939) \rev{and for ViT-B/32 images were normalised using ImageNet mean (0.485, 0.456, 0.406) and standard deviation (0.229, 0.224, 0.225)}. Unless otherwise specified, the weights of all of the models were transferred from a model trained on the ImageNet dataset.

\noindent\textbf{Model ensembling:} in the third step of the proposed pipeline, we combined probability maps generated by individual models using ensembling. Specifically, several probability maps were combined by averaging the probabilities returned by the individual models. While using ensembling to improve performance is a common practice, in the histopathological image recognition task, in addition to combining the predictions made by different architectures of neural networks, we also propose combining models trained on different tissue magnifications. We take advantage of the fact that, while the model trained on a higher magnification will have to make predictions for a larger number of patches to encompass the same region as the model trained on a lower magnification, spatially, they correspond to the same scan region. As a result, we simply rescale probability maps generated by higher magnification models to a common dimensionality, after which we can once again combine their predictions by map averaging.

\noindent\textbf{Scan-level diagnosis:} finally, to translate the patch-level probability maps generated in previous steps into a single scan-level diagnosis we considered an approach of thresholding the ratio of scan patches that were classified as cancerous to the overall number of valid tissue patches. Specifically, we used a simple decision-making rule based on two parameters, lower threshold $T_L$, and upper threshold $T_U$, based on the percentage of valid tissue patches (that is, all patches excluding BG class and scan background excluded during the initial segmentation step) of a given scan $p_c$, gave a diagnosis 'non-cancerous' if $p_c \leq T_L$, 'cancerous' if $p_c \geq T_U$, and abstained from making the decision if $T_L < p_c < T_U$. Such decision rule can be interpreted as a simple abstaining classification algorithm, which optimizes a multi-objective criterion: on the one hand, we wish to achieve as high diagnosis accuracy as possible, while on the other, we wish to abstain from giving the diagnosis in the least possible number of cases. Both criteria are clearly opposing, since by reducing the width of the range in which we abstain from making the prediction, we decrease the chance of error, and vice versa. It is also unclear what cost should be assigned to an incorrect diagnosis and to abstaining from making the decision, making the problem ambiguous. Because of that, instead of presenting a single result of a chosen model, we examined the trade-off associated with choosing different values of $T_L$ and $T_U$. 

\section*{Results}
\label{sec:expstudy}


\subsection*{Comparison of different convolutional neural networks in the patch recognition task}
\label{subsec:cnn}

In the first stage of the conducted experimental study, we evaluated the impact of the choice of a convolutional neural network architecture on the performance of the patch recognition task. During this comparison, we used DiagSet-A.1, which is the initial patch recognition dataset, prior to adding the data acquired in the weakly supervised manner. We considered three different classification scenarios. First of all, a binary setting, in which data was divided into two classes based on the associated label: either a non-cancerous, containing tissue background (T), healthy tissue (N) or artifacts (A), or cancerous, which was labeled with any of the Gleason scores (R1-5) by the histopathologist. In the second setting, we considered the Gleason scores equal or higher to 3 separately, treating them as individual classes, and merged the remaining labels (T, N, A, R1 and R2) into a single class; this partitioning resulted in a total of 4 classes. Finally, in the third setting, all of the labels were considered separately, producing a classification task consisting of 8 classes. From a practical standpoint the primary consideration is whether we intend to predict a specific Gleason score or are satisfied with a simple binary (cancerous or non-cancerous) diagnosis; differentiation between the remaining classes, that is tissue background, healthy tissue, and artifacts is, however, informative from the point of view of understanding the systems behavior and its inter-class errors. The choice of treating the lowest Gleason scores, equal to 1 or 2, as either cancerous or non-cancerous is also debatable due to the fact that, because of high similarity to the healthy tissue, they are not recommended for usage by medical practitioners. Finally, it is worth noting that during the conducted experiments we did not observe a difference between the performance of models trained to discriminate all of the available classes, with predictions merged after the fact by summation of the individual class probabilities, and the performance of a model trained from a get-go on a merged set of classes. This suggests that training the classification network in a multi-class fashion is a suitable approach regardless of whether we intend to perform more detailed discrimination than the binary variant.

The results are presented in Table~\ref{table:patch-recognition-all}. As can be seen, in the binary setting the best performance was achieved by the VGG19 architecture, which was able to obtain an above 90-percent classification accuracy for all tissue magnifications. Interestingly, VGG architectures outperformed some of the more recent models known to achieve a better classification accuracy in a natural image recognition task, namely ResNet, Inception \rev{and ViT-B/32. The transformer architecture achieved the worst observed result at the highest magnification. This confirms already existing studies \cite{LiVT, Li_Duggal_Chau_2023} about ViTs heavy degradation when training on imbalanced datasets.} Based on the results achieved in the binary setting two \rev{convolutional} neural architectures, namely VGG19 and ResNet50 \rev{and single transformer architecture, ViT-B/32}, were selected for comparison in the multi-class setting. Because class imbalance becomes more significant outside the binary variant, in addition to the traditional classification accuracy, we also present the average accuracy (AvAcc). 
Note that AvAcc was not recorded for the binary setting. However, because the number of observations from both classes was roughly the same, the accuracy and AvAcc are highly correlated. As can be seen, in both multi-class variants, VGG19 achieved better classification accuracy, with disproportion between the results increasing with the number of classes. However, ResNet50 achieved better performance with respect to AvAcc for all of the magnifications in the four-class setting, and for a single magnification in the eight-class setting, indicating lower bias of the network toward the majority class. \rev{In comparison to both CNN architectures, ViT-B/32 is characterized by the biggest difference between accuracy and AvAcc, indicating problems with generalization and existing classification bias. }
In general, while VGG19 achieved the best performance on the most represented classes, T, N, and R4, ResNet50 scored better on less represented classes, such as A and Gleason scores other than 4. This indicates a diversity of predictions made by different neural architectures, suggesting the suitability of ensembling techniques.


\begin{table}
\small
\caption{Accuracy and average accuracy obtained using the specific architectures of the convolutional neural networks \rev{and vision transformer} in the multi-class setting: either a) 2-class setting, with discrimination between any of the cancerous tissue and any of the other classes, b) 4-class setting, with Gleason scores 3-5 treated as individual classes, or c) with all classes being considered separately (no class merging).}
\label{table:patch-recognition-all}
\centering
\resizebox{\textwidth}{!}{
\begin{tabular}{llllllllllllll}
\toprule
& & \multicolumn{6}{c}{\textbf{2-class setting}} & \multicolumn{3}{c}{\textbf{4-class setting}} & \multicolumn{3}{c}{\textbf{8-class setting}} \\
\cmidrule(r){3-8} \cmidrule(r){9-11} \cmidrule(r){12-14}
\textbf{Met.} & \textbf{Mag.} & \textbf{AlexNet} & \textbf{VGG16} & \textbf{VGG19} & \textbf{ResNet50} & \textbf{InceptionV3} & \rev{\textbf{ViT-B/32}} & \textbf{VGG19} & \textbf{ResNet50} & \rev{\textbf{ViT-B/32}} & \textbf{VGG19} & \textbf{ResNet50} & \rev{\textbf{ViT-B/32}} \\
\midrule
Acc & $40\times$ & 89.72 & 92.53 & \textbf{92.89} & 89.34 & 86.07 & \rev{78.96} &\textbf{85.81} & 78.52 & \rev{81.04} & \textbf{62.89} & 39.60  & \rev{61.27} \\
 & $20\times$ & 90.19 & 91.95 & \textbf{92.39} & 85.54 & 88.56 & \rev{90.04} & \textbf{85.26} & 73.54 & \rev{81.46} & \textbf{62.51} & 42.97 & \rev{61.82} \\
 & $10\times$ & 88.67 & 91.58 & \textbf{91.86} & 86.92 & 85.61 & \rev{89.40} & \textbf{83.89} & 72.05 & \rev{76.48} & \textbf{63.21} & 42.40 & \rev{62.40} \\
 & $5\times$ & 84.32 & 90.03 & \textbf{90.24} & 86.19 & 86.01 & \rev{86.57} & \textbf{81.62} & 73.56 & \rev{77.64} & \textbf{63.44} & 43.94  & \rev{60.63} \\
\midrule
AvAcc & $40\times$ & - & - & - & - & - & - & 51.47 & \textbf{51.62} & \rev{43.69} &  \textbf{39.16} & 35.02 & \rev{29.08} \\
 & $20\times$ & - & - & - & - & - & - & 55.06 & \textbf{57.87} & \rev{43.23} & \textbf{38.89} & 36.95 & \rev{33.15} \\
 & $10\times$ & - & - & - & - & - & - & 55.96 & \textbf{58.36} & \rev{37.49} & \textbf{37.60} & 35.40 & \rev{32.05} \\
 & $5\times$ & - & - & - & - & - & - & 55.36 & \textbf{59.20} & \rev{46.38} & 35.15 & \textbf{37.28} & \rev{29.26} \\
\bottomrule
\end{tabular}}
\end{table}

\subsection*{Building ensembles of convolutional neural networks}

A common strategy for improving the performance of machine learning models is building classifier ensembles. This approach combines outputs of several underlying models to form a single, combined prediction. An essential requirement for achieving a performance improvement when using ensembles is ensuring sufficient diversity of predictions of the underlying models. In the context of computer vision and deep learning, this is often achieved by combining several different architectures of convolutional neural networks. 
However, in addition to achieving diversity by combining different neural architectures, histopathological tissue classification, in principle, enables the ensembling of models trained on different tissue magnification. To empirically test the practical usefulness of such an approach, we conducted an experiment, in which we combined various convolutional neural networks, differing for both the neural architecture, as well as the considered tissue magnification. Specifically, we once again considered AlexNet, VGG16, VGG19, ResNet50 and InceptionV3 networks, each trained on either $40\times$, $20\times$, $10\times$ or $5\times$ magnification. Afterward, the probabilities returned by the individual models were combined via averaging, either at the model or magnification level, or both. The task considered was binary classification, which is a discrimination between the cancerous and non-cancerous patches, and the performance was evaluated using standard classification accuracy. The results are presented in Table~\ref{table:ensemble-mag}. As can be seen, both ensembling strategies produced results better than the individual models: for the family of models trained on a single magnification, ensembling all five architectures produced the best results in the case of every magnification. Similarly, the performance of specific models was also improved in every case by combining different magnifications: depending on the model, either combining all of the available magnifications, or all but $5\times$ magnification (on which models tended to achieve the worst performance), produced the best results. Finally, the best overall performance was achieved by combining both modes of ensembling, that is for an ensemble of all available architectures trained on $40\times$, $20\times$, and $10\times$ magnifications, indicating that ensembling on a magnification and model level are complementary.


It is also worth mentioning that in the patch recognition task predictions made for spatially nearby patches are not uncorrelated: since cancerous tissue tends to form larger clusters containing multiple tissue patches, the presence of non-cancerous neighbors decreases the probability of a given patch being cancerous itself. Because of that, an alternative to the traditional ensembling of predictions produced by multiple models can be correcting the predictions of a single model based on the predictions made for neighboring patches. A conceptually simple implementation of this idea is applying median filtering to post-process the prediction map produced by a given model, an approach aimed at eliminating individual outliers that do not form larger clusters. To empirically evaluate the usefulness of such an approach, we repeated the previously described ensembling experiment, this time post-processing the prediction maps produced by every individual model or ensemble via median filtering with kernel size $k = 3$. The results are presented in Table~\ref{table:ensemble-mag}. As can be seen, applying median filtering allowed us to achieve slightly better performance for the ensemble consisting of all of the considered neural architectures and $40\times$, $20\times$ and $10\times$ magnifications. Furthermore, perhaps more importantly, it allowed us to achieve the same performance for an ensemble consisting of a significantly lower number of models, namely two VGG16 networks trained at $40\times$, and $20\times$ magnifications, significantly reducing the computational overhead associated with training and interference. However, it is worth noting that to enable the use of median filtering, the classification problem had to be binarized, and extending it to the multi-class setting would require further extensions. Nevertheless, overall, the observed results indicate the usefulness of classifier ensembles in the patch recognition task. In particular, both ensembling models trained on different data magnifications, as well as spatially correcting the predictions of the model, seem to offer a suitable alternative to traditional ensembling across different model architectures.

\begin{table*}
\small
\caption{Results achieved by an ensemble of different neural architectures and/or different magnification factors, either without median filtering (top) or with it (bottom).}
\label{table:ensemble-mag}
\centering
\begin{tabular}{llllllllll}
\toprule
\textbf{Model} & \textbf{$40\times$} & \textbf{$20\times$} & \textbf{$10\times$} & \textbf{$5\times$} & \textbf{$40{+}20\times$} & \textbf{$40{+}20{+}10\times$} & \textbf{$40{+}20{+}10{+}5\times$} & \textbf{$20{+}10\times$} & \textbf{$20{+}10{+}5\times$} \\
\midrule
AlexNet & 89.66 & 91.07 & 89.51 & 83.98 & 91.98 & \textbf{92.67} & 92.28 & 91.86 & 90.94 \\
VGG16 & 92.47 & 93.14 & 91.57 & 88.15 & 94.03 & \textbf{94.26} & 93.97 & 93.49 & 92.55 \\
VGG19 & 92.32 & 93.16 & 91.67 & 87.92 & 94.01 & \textbf{94.15} & 93.87 & 93.55 & 92.58 \\
ResNet50 & 88.29 & 87.20 & 88.14 & 86.20 & 90.02 & 91.86 & \textbf{93.14} & 91.53 & 92.00 \\
InceptionV3 & 86.09 & 90.50 & 86.45 & 87.12 & 90.60 & 92.96 & \textbf{93.54} & 92.42 & 92.29 \\
5 CNN ensemble & 92.99 & 93.74 & 92.28 & 88.48 & 94.22 & \textbf{94.51} & 94.31 & 93.86 & 93.13 \\
\midrule
AlexNet & 93.20 & 92.97 & 90.02 & 84.16 & \textbf{93.76} & 93.68 & 93.00 & 92.82 & 91.50 \\
VGG16 & 94.26 & 94.19 & 91.92 & 88.25 & \textbf{94.67} & 94.66 & 94.19 & 93.96 & 92.81 \\
VGG19 & 94.28 & 94.18 & 92.00 & 88.00 & \textbf{94.66} & 94.54 & 94.11 & 94.00 & 92.80 \\
ResNet50 & 89.75 & 89.08 & 88.65 & 86.37 & 91.09 & 92.29 & \textbf{93.49} & 92.35 & 92.47 \\
InceptionV3 & 85.69 & 92.37 & 87.24 & 87.28 & 90.83 & 93.53 & \textbf{93.91} & 93.41 & 92.77 \\
5 CNN ensemble & 93.92 & 94.55 & 92.58 & 88.56 & 94.59 & \textbf{94.67} & 94.53 & 94.23 & 93.34 \\
\bottomrule
\end{tabular}
\end{table*}


\subsection*{Evaluation of the final model in the patch recognition task}
\label{sec:final-patch-rec}

To avoid overfitting all of the results presented up to this point were obtained on the validation partition of the dataset. However, to serve as a reference point for further studies, we also evaluated the performance of the final model on the test partition of the data. Additionally, compared to the previous experiments, which used the DiagSet-A.1 version of the dataset, the final performance was reported after training on the DiagSet-A.2, which contained a total of 346 annotated WSI. We examined the performance of an ensemble consisting of VGG19 models trained separately on $40\times$, $20\times$, $10\times$ and $5\times$ magnifications. We evaluated five different class settings, each with a different level of granularity. Class settings 1 (S1) and 2 (S2) correspond to the binary classification, with Gleason grades 1 and 2 treated as either cancerous (S1) or non-cancerous (S2) tissue. In setting 3 (S3), additional discrimination between Gleason grades 3-5 was introduced, and grades 1-2 were added to the non-cancerous class group, leading to 4 total classes. In setting 4 (S4), every Gleason grade was assigned a separate class, leading to 6 total classes. Finally, in setting 5 (S5), we also introduced the discrimination between the non-cancerous classes, treating each separately, leading to 9 different classes.



The summary of the results is presented in Table~\ref{table:patch-recognition-final}. 
As can be seen, in all but the last setting, in which a discrimination between non-cancerous tissue classes was introduced, an above 90 percent classification accuracy was observed, with the best performance observed in the first binary setting, for which 94.58 accuracy and 94.70 AvAcc were achieved. The differences between settings 1 and 2 were generally negligible due to a low percentage of scans containing Gleason grades 1 and 2. It is worth noting that the majority of inter-errors were observed between the healthy tissue (N) and tissue background (T), likely because non-cancerous regions were labeled in less detail, introducing the most significant degree of label noise between those two classes; and Gleason grades 3-5, the discrimination between which was in general more subjective than the discrimination between cancerous and non-cancerous tissue. It is also worth noting that the model in almost no case was able to recognize Gleason grades 1 and 2, labeling them as either healthy tissue or Gleason 3-4 instead. A similar behavior was also displayed for the artifacts, which in most classes were categorized as healthy tissue. 

\begin{table}
\small
\caption{Final performance of the selected model on test partition in patch recognition task, in one of the 5 class settings.}
\label{table:patch-recognition-final}
\centering
\begin{tabular}{llllll}
\toprule
\textbf{Metric} & \textbf{S1} & \textbf{S2} & \textbf{S3} & \textbf{S4} & \textbf{S5} \\
\midrule
Acc & 94.58 & 94.53 & 92.70 & 92.00 & 69.36 \\
AvAcc & 94.70 & 94.41 & 62.91 & 42.17 & 44.92 \\
\bottomrule
\end{tabular}
\end{table}


\subsection*{Evaluation of the capabilities of a complete system in the scan-level diagnosis}
\label{sec:scan-level-diagnosis}

\begin{figure}
\centering
\includegraphics[width=0.4\linewidth]{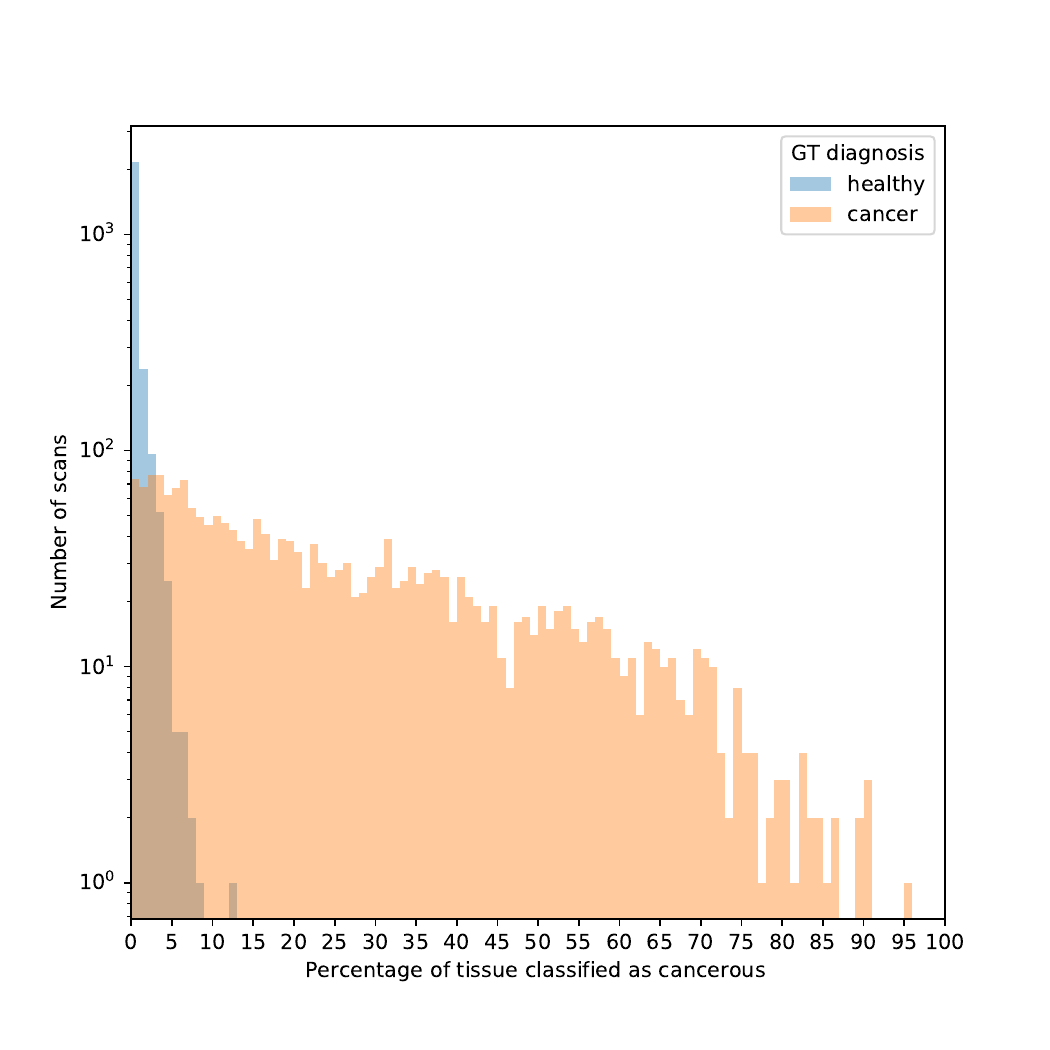}
\includegraphics[width=0.4\linewidth]{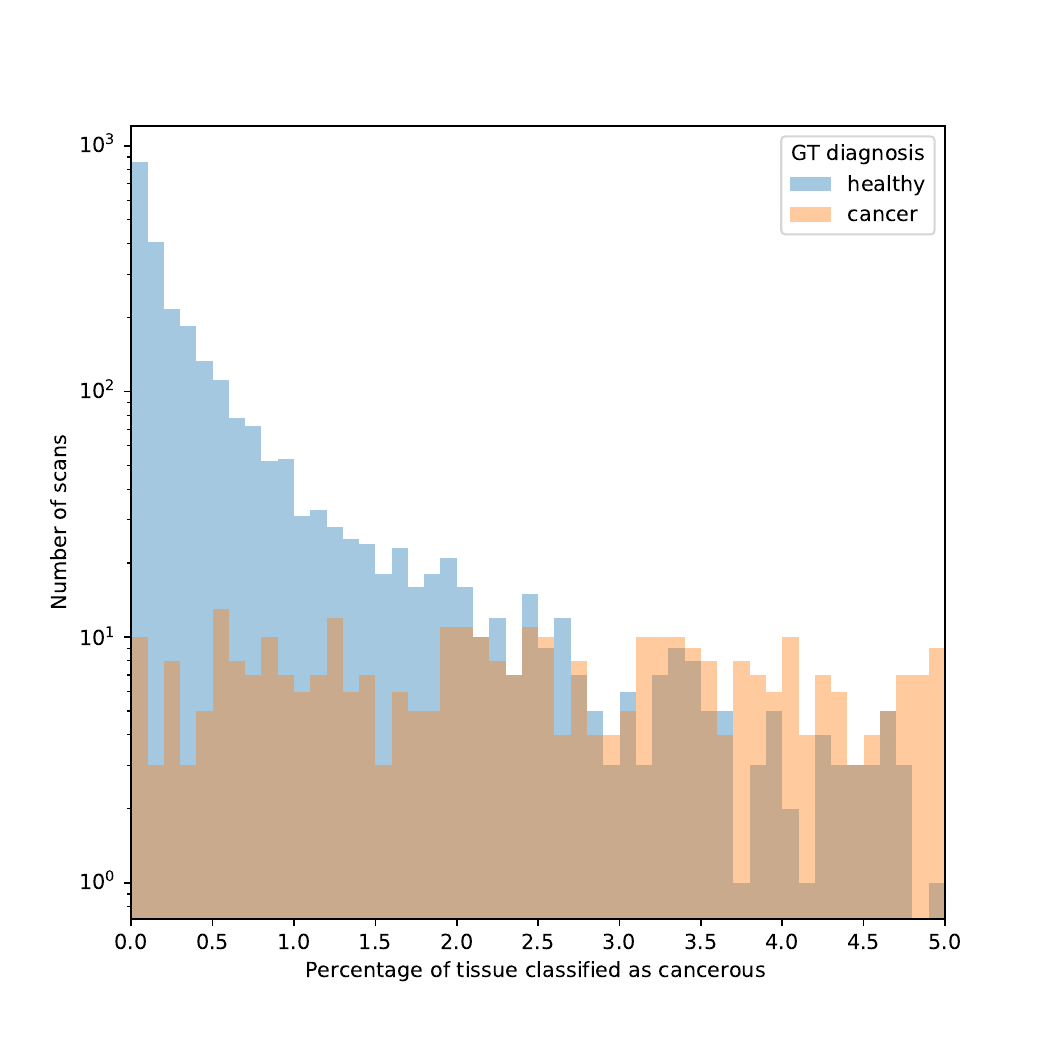}
\caption{Relation between the percentage of tissue classified as cancerous and ground truth diagnosis given by the histopathologist: in the full range (left) and magnified in the 0-5\% range (right). Note that the logarithmic scale was used.}
\label{fig:diag-acc}
\end{figure}

In addition to evaluating the performance of the proposed methodology in the patch recognition task, we also considered the problem of predicting a final diagnosis for a complete WSI scan. We focused on \textit{the binary diagnosis}, that is, classifying the scan as either cancerous or non-cancerous, with the possibility of abstaining from deciding uncertain cases, implying that either further confirmation by a human histopathologist or scheduling IHC is necessary. We analyzed the DiagSet-B dataset, which consisted of 4,675 WSI scans with an associated binarized diagnosis made by a human histopathologist. We began by examining the percentage distribution of tissue patches classified as cancerous by the ensemble of convolutional neural networks. This distribution is presented in Figure~\ref{fig:diag-acc}. As can be seen, there is a clear percentage cut-off, after which every scan in the considered dataset was diagnosed as cancerous by a human histopathologist, and about 13\% of tissue was classified as cancerous by the considered ensemble. However, such unambiguousness was not observed in the lower range of the detected cancerous tissue percentage. Still, regions in which one of the diagnoses dominated can be distinguished at around $(0, 2)$ and $(5, 13)$, indicating some discriminatory capabilities. Finally, neither of the diagnoses dominated in the $(2, 5)$ region, clearly indicating the need of abstaining from the diagnoses.

Based on the observed distribution, we considered an approach for scan-level diagnosis, which relied on the statistical distribution of predictions made for a given scan in the patch recognition task. To reiterate, the proposed approach was based on a simple decision-making rule using two parameters, lower threshold $T_L$, and upper threshold $T_U$, based on the percentage of valid tissue patches (that is, all patches excluding BG class and scan background excluded during the initial segmentation step) of a given scan $p_c$, gave a diagnosis 'non-cancerous' if $p_c \leq T_L$, 'cancerous' if $p_c \geq T_U$, and abstained from making the decision if $T_L < p_c < T_U$.

\begin{figure*}
\centering
\includegraphics[width=0.49\linewidth]{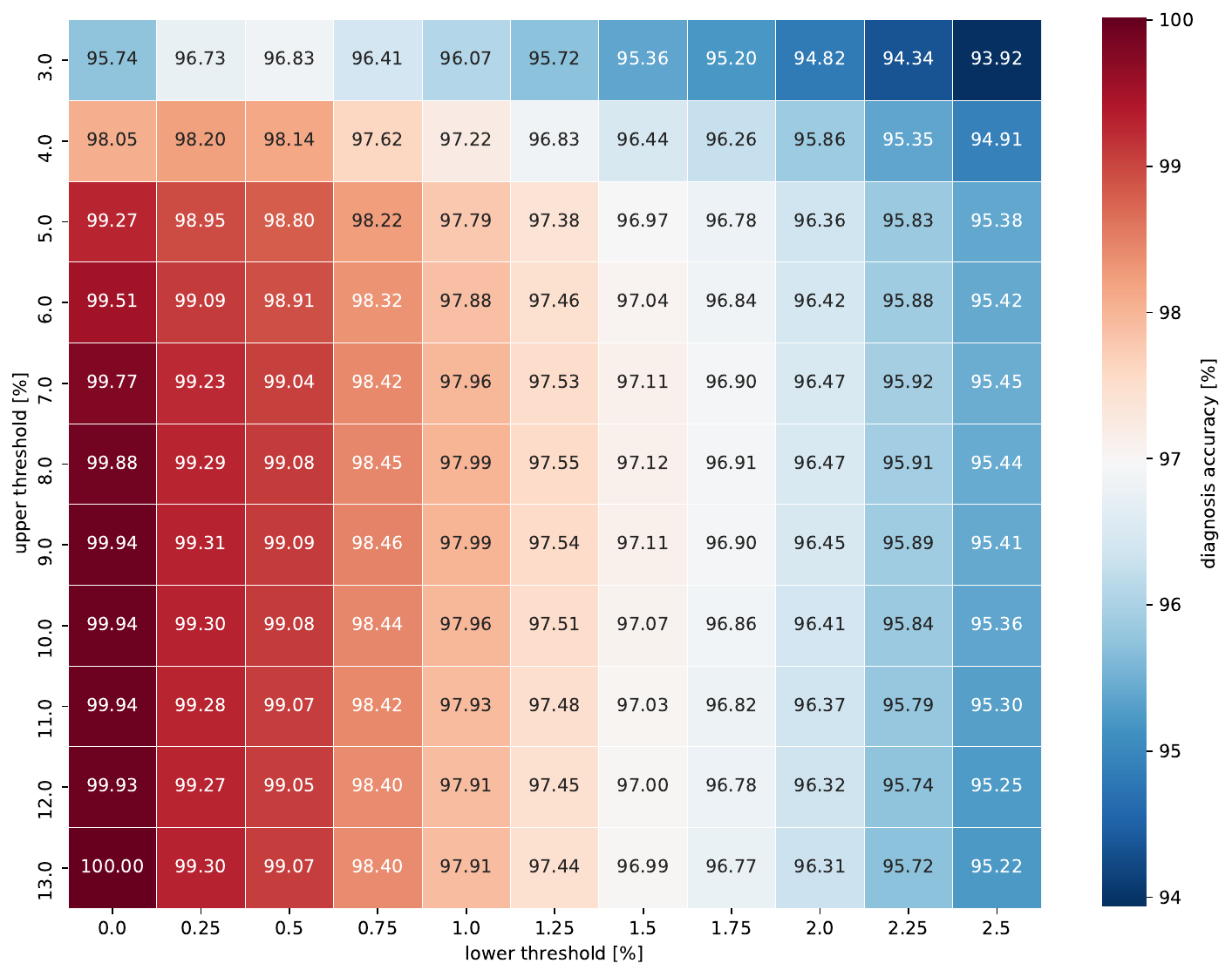}
\includegraphics[width=0.49\linewidth]{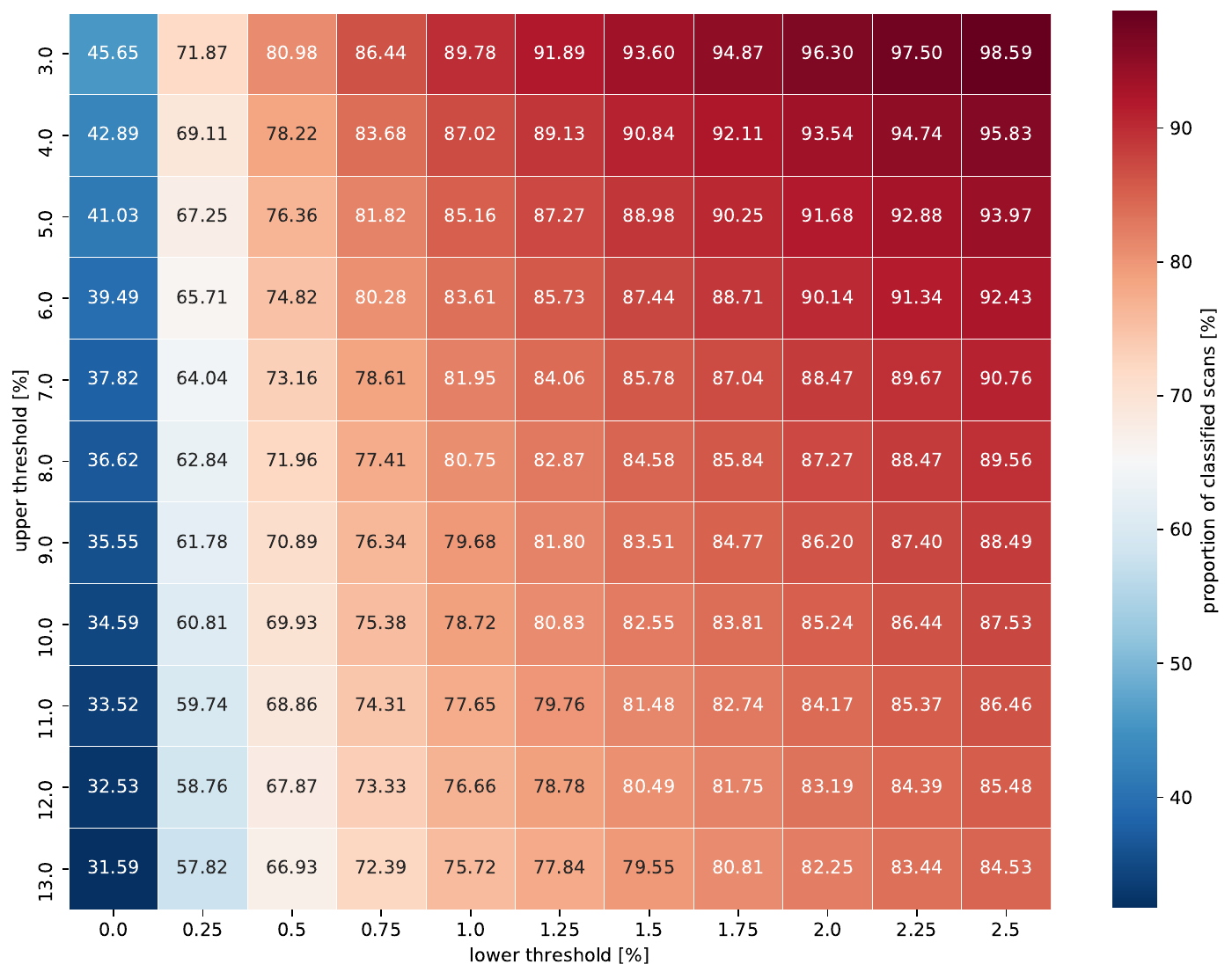}
\caption{The impact of setting the lower and upper threshold used to give a final diagnosis on the diagnosis accuracy (left) and the proportion of classified scans (right).}
\label{fig:thresholding}
\end{figure*}

Two matrices, containing the impact of the parameters on both the accuracy and the proportion of scans for which a diagnosis was given, are presented in Figure~\ref{fig:thresholding}. As can be seen, by properly setting the threshold values, we are able to achieve a desired trade-off between the quality and the quantity of the predictions. For example, setting the lower threshold at 0.5\% and the upper threshold at 7\%, which corresponds to the case in which we diagnose a scan as cancerous if the percentage of patches classified as cancerous is lower than 0.5\%, as non-cancerous if it is higher than 7\%, and we abstain from making the decision in the remaining cases, we achieve a 99.04\% diagnosis accuracy, at the same time processing 73.16\% of scans, or in other words abstaining from making the decision for 26.84\% of the scans. On the other hand, in the extremes, we are able to achieve either a perfect diagnosis accuracy on the considered dataset, in which case, however, we only process 31.59\% of the scans, or process almost all of the scans, that is 98.59\% of them, at the same time achieving a lower diagnosis accuracy of 93.92\%. The observed results indicate that while using a simple decision rule, in which we base the diagnosis on the percentage of tissue classified as cancerous, is not sufficient to achieve a perfect performance for all of the cases, by properly setting the range in which we abstain from giving the diagnosis we can significantly improve the performance at the cost of the number of processed scans. In particular, the scans for which a high percentage of tissue was classified as cancerous led to high certainty of correct diagnosis, whereas more ambiguity was present at the opposite end of the spectrum, at which even a minimal number of cancerous patches could have indicated the actual presence of cancer. As a result, this suggests the system's suitability for the initial screening, during which we give the diagnosis only for the scans we deem least ambiguous.

\subsection*{Comparison with human histopathologists}

To assess the coherence of the medical diagnosis of prostate cancer an experiment with 9 volunteered physicians, experts in histopathology, has been conducted. They have been presented with the 46 WSI of prostate tissue, some containing cancerous changes. The same has been tried with the DCNN diagnosis rule described earlier. The results of this experiment are presented in Table~\ref{table:doctors-experiment}. Each D\textit{i} column contains responses of the \textit{i}-th expert, whereas DCNN corresponds to the responses of the proposed system. The "C Tissue \%" column reflects the percentage of the cancerous tissues in each WSI scan as returned by the ensemble of deep classifiers.



\begin{table*}
\caption{Results of the experiment of the prostate cancer diagnosis experiment performed by the nine histopathology experts D\textit{i}, DCNN and the three statistical hypothesis tests. "C Tissue \%" contains a percentage of cancerous tissues detected by the ensemble of deep classifiers.}
\label{table:doctors-experiment}
\centering
\begin{tabular}{llllllllllrl}
\toprule
WSI No. & D1  & D2  & D3  & D4  & D5  & D6  & D7  & D8  & D9  & C Tissue \%   & DCNN \\
\midrule
1   & C   & C   & C   & C   & C   & C   & C   & C   & C   & 13.35 & C  \\
2   & C   & C   & C   & IHC & C   & C   & C   & C   & C   & 21.77 & C  \\
3   & C   & C   & C   & C   & C   & C   & C   & C   & C   & 13.78 & C  \\
4   & C   & C   & C   & C   & C   & IHC & C   & C   & C   & 10.90 & C  \\
5   & C   & C   & C   & C   & C   & C   & C   & C   & C   & 35.16 & C  \\
6   & C   & C   & C   & C   & C   & C   & C   & C   & C   & 34.80 & C  \\
7   & C   & C   & C   & C   & C   & C   & C   & C   & C   & 46.16 & C  \\
8   & C   & C   & C   & C   & C   & C   & C   & C   & C   & 46.78 & C  \\
9   & C   & C   & C   & C   & C   & C   & C   & C   & C   & 24.83 & C  \\
10  & C   & IHC & NC  & C   & NC  & NC  & NC  & IHC & IHC & 1.87  & IHC  \\
11  & C   & C   & C   & C   & C   & C   & C   & C   & C   & 33.22  & C  \\
12  & C   & C   & C   & C   & C   & C   & C   & C   & C   & 3.90  & IHC  \\
13  & C   & C   & IHC & IHC & NC  & IHC & IHC & C   & C   & 0.33  & NC  \\
14  & NC  & NC  & NC  & NC  & NC  & NC  & NC  & IHC & NC  & 0.00     & NC  \\
15  & C   & NC  & IHC & IHC & IHC & NC  & IHC & IHC & IHC & 0.15  & NC  \\
16  & NC  & NC  & NC  & NC  & NC  & NC  & NC  & NC  & NC  & 0.06  & NC  \\
17  & C   & C   & C   & C   & C   & C   & C   & C   & C   & 25.04 & C  \\
18  & C   & C   & C   & C   & C   & C   & C   & C   & C   & 46.00 & C  \\
19  & IHC & IHC & NC  & IHC & NC  & IHC & NC  & IHC & IHC & 2.16  & IHC   \\
20  & IHC & IHC & NC  & IHC & NC  & NC  & NC  & NC  & IHC & 0.09  & NC  \\
21  & IHC & IHC & NC  & IHC & NC  & IHC & IHC & IHC & IHC & 1.33  & IHC  \\
22  & C   & C   & C   & C   & C   & C   & C   & C   & C   & 8.51  & C   \\
23  & C   & C   & C   & C   & C   & C   & C   & C   & C   & 69.31  & C   \\
24  & IHC & C   & C   & C   & C   & C   & C   & C   & C   & 2.55  & IHC   \\
25  & IHC & C   & C   & IHC & C   & C   & C   & C   & C   & 2.46  & IHC   \\
26  & IHC & C   & C   & C   & C   & C   & C   & C   & C   & 3.49  & IHC   \\
27  & C   & C   & C   & C   & C   & C   & C   & C   & C   & 5.87   & IHC    \\
28  & C   & C   & C   & C   & C   & C   & C   & IHC & C   & 69.49 & C   \\
29  & C   & C   & C   & C   & C   & C   & C   & C   & C   & 54.05 & C    \\
30  & IHC & C   & IHC & C   & C   & C   & C   & C   & C   & 8.76  & C  \\
31  & C   & C   & C   & C   & C   & C   & C   & C   & C   & 13.38 & C   \\
32  & C   & C   & C   & C   & C   & C   & C   & C   & C   & 10.02 & C    \\
33  & IHC & IHC & NC  & IHC & NC  & NC  & NC  & IHC & NC  & 6.42  & IHC  \\
34  & C   & C   & C   & C   & C   & C   & C   & C   & C   & 29.68 & C     \\
35  & C   & C   & C   & C   & C   & C   & C   & C   & C   & 55.96   & C     \\
36  & C   & C   & C   & C   & C   & C   & C   & C   & C   & 37.92 & C \\
37  & C   & C   & C   & C   & C   & C   & C   & C   & C   & 10.21 & C   \\
38  & C   & C   & C   & C   & C   & C   & C   & C   & C   & 27.60 & C   \\
39  & C   & C   & C   & C   & C   & C   & C   & C   & C   & 14.54 & C  \\
40  & C   & C   & C   & C   & C   & C   & C   & C   & C   & 32.48  & C   \\
41  & NC  & NC  & NC  & NC  & NC  & NC  & NC  & NC  & NC  & 0.39  & NC   \\
42  & NC  & NC  & NC  & IHC & NC  & NC  & NC  & IHC & IHC & 5.13  & IHC  \\
43  & C   & C   & C   & C   & C   & C   & C   & C   & C   & 21.11 & C   \\
44  & IHC & NC  & IHC & NC  & IHC & IHC & NC  & IHC & IHC & 0.13  & NC   \\
45  & NC  & NC  & NC  & NC  & NC  & NC  & NC  & NC  & NC  & 0.49  & NC    \\
46  & NC  & NC  & NC  & NC  & NC  & NC  & NC  & NC  & IHC & 0.40  & NC  \\
\bottomrule
\end{tabular}
\end{table*}

Finally, correlations of the answers shown in Table~\ref{table:doctors-experiment} have been statistically verified with the help of the Spearman non-parametric test \cite{Sheskin2007}. Each pair of responses has been verified, including the responses of a human vs. human, as well as human vs. the proposed DCNN rule. The results of these tests are presented in Table~\ref{table:Spearman-experiment}. The correlation between the DCNN method and expert histopathologists is in the range from 0.75 up to 0.83. These results do not significantly differ from the correlations obtained in the group of human experts, in which the lowest correlation was 0.64.  

\begin{table*}
\small
\caption{Non parametric Spearman correlation coefficients between the 9 histopathologists and DCNN on examination of 46 WSI with various stadia of the prostate cancer.}
\label{table:Spearman-experiment}
\centering
\begin{tabular}{lllllllllll}
\toprule
     & D1   & D2   & D3   & D4   & D5   & D6   & D7   & D8   & D9   \\
\midrule
D1   & 1.00 & - & - & - & - & - & - & - & -  \\
D2   & 0.73 & 1.00 & - & - & - & - & - & - & -  \\
D3   & 0.74 & 0.92 & 1.00 & - & - & - & - & - & -  \\
D4   & 0.78 & 0.82 & 0.79 & 1.00 & - & - & - & - & -  \\
D5   & 0.68 & 0.92 & 0.97 & 0.82 & 1.00 & - & - & - & -  \\
D6   & 0.64 & 0.93 & 0.93 & 0.79 & 0.95 & 1.00 & - & - & -  \\
D7   & 0.70 & 0.95 & 0.96 & 0.83 & 0.99 & 0.96 & 1.00 & - & -  \\
D8   & 0.70 & 0.94 & 0.89 & 0.77 & 0.89 & 0.89 & 0.92 & 1.00 & -  \\
D9   & 0.73 & 0.98 & 0.93 & 0.81 & 0.93 & 0.93 & 0.96 & 0.94 & 1.00  \\
\midrule
DCNN & 0.75 & 0.79 & 0.78 & 0.83 & 0.82 & 0.79 & 0.82 & 0.74 & 0.78 &  \\
\bottomrule
\end{tabular}
\end{table*}


\section*{Conclusions}
\label{sec:concl}

In this paper, we present the DiagSet dataset containing whole slide images of the prostate cancer tissues annotated and labeled by the professional histopathology doctors following the Gleason scale. This dataset made possible training and testing of several deep convolutional neural network architectures, such as AlexNet, VGG16, VGG19, ResNet50, and InceptionV3, \rev{as well as vision transformer architecture ViT-B/32}, in various configurations, data imbalance settings, and different image magnifications. From these, the best-performing configurations were selected to form an ensemble of deep neural classifiers, for which the obtained best accuracy is 94.6{\%} in the binary cancer vs. no-cancer classification setup. 
Based on the patch-level predictions, a simple strategy for whole-slide diagnosis based on thresholding was devised.
The obtained results of machine-based diagnosis have been verified in the experiment involving histopathology doctors, showing high statistical agreement.
We believe that the methods presented here, as well as the fully annotated DiagSet - available from the Internet - will be very useful to other researchers.
However, we plan to continue working in this direction. Among other things, future research will focus on examining our ensembles with other prostate cancer datasets, as well as those connected to DiagSet. Future work will also include new architectures based on attention mechanisms in ViT structures.


\section*{Data availability}

The datasets generated and analyzed during the current study are available at \url{https://github.com/michalkoziarski/DiagSet}.

\bibliography{refs}

\section*{Acknowledgements}

This research was co-financed by the European Regional Development Fund in the Intelligent Development 2014-2020 Programme, within the grant “The system of automatic analysis and recognition of histopathological images” supported by the National Center for Research and Development: grant no. POIR.01.01.01-00-0861/16-00, and Diagnostyka Consilio.

\section*{Author contributions statement} 

B.O., Z.A., B.C., B.K. and M.K. designed the grading and data labeling protocol. B.O., Z.A. and M.Ż. annotated the scans. M.K. supervised labeling process, developed the classification pipeline and performed all experiments except ViT. P.N. performed ViT experiments. M.K. and B.C. performed data analysis. M.K. and A.B. implemented the approach. B.C. and B.K. supervised the research. P.W. developed the dataset website. B.C., B.K., P.S., M.K. and J.S. conceptualised the project. M.K. and B.C. wrote the manuscript. All authors reviewed the manuscript.

\section*{Additional information} 

\textbf{Competing interests}: The authors declare no competing interests.


\end{document}